\documentclass[aps,twocolumn,pra,superscriptaddress,showpacs]{revtex4}
\usepackage{mathrsfs}
\usepackage{amssymb}
\usepackage{amsmath}
\usepackage{graphicx}
\usepackage{epsfig}
\usepackage{txfonts}
\usepackage{subfigure}
\usepackage{amsfonts}
\usepackage{color}
\usepackage[colorlinks,citecolor=blue]{hyperref}
\usepackage{bm}
\begin{document}

\title{Quantum sensing of temperature close to absolute zero in a Bose-Einstein condensate}
\author{Ji-Bing Yuan}
\affiliation{College of Physics and Electronic Engineering, Hengyang Normal University, and Key Laboratory of Opto-electronic Control and Detection Technology of University of Hunan Province, Hengyang 421002, China}
\author{Bo Zhang}
\affiliation{College of Physics and Electronic Engineering, Hengyang Normal University, and Key Laboratory of Opto-electronic Control and Detection Technology of University of Hunan Province, Hengyang 421002, China}
\author{Ya-Ju Song}
\affiliation{College of Physics and Electronic Engineering, Hengyang Normal University, and Key Laboratory of Opto-electronic Control and Detection Technology of University of Hunan Province, Hengyang 421002, China}
\author{Shi-Qing Tang}
\affiliation{College of Physics and Electronic Engineering, Hengyang Normal University, and Key Laboratory of Opto-electronic Control and Detection Technology of University of Hunan Province, Hengyang 421002, China}
\author{Xin-Wen Wang}
\affiliation{College of Physics and Electronic Engineering, Hengyang Normal University, and Key Laboratory of Opto-electronic Control and Detection Technology of University of Hunan Province, Hengyang 421002, China}
\author{Le-Man Kuang\footnote{Author to whom any correspondence should be
addressed. Email: lmkuang@hunnu.edu.cn}}

\affiliation{Key Laboratory of Low-Dimensional Quantum Structures
and Quantum Control of Ministry of Education,  and Department of
Physics, Hunan Normal University, Changsha 410081, China}
\affiliation{Synergetic Innovation Academy for Quantum Science and Technology, Zhengzhou University of Light Industry, Zhengzhou 450002, China}
\date{\today}

\begin{abstract}
We propose a theoretical scheme for quantum sensing of temperature close to absolute zero in a quasi-one-dimensional Bose-Einstein condensate (BEC). In our scheme, a single-atom impurity qubit is used as a temperature sensor. We investigate the sensitivity of the single-atom  sensor in estimating the temperature of the BEC. We demonstrate that the sensitivity of the temperature sensor can saturate the quantum Cram\'{e}r-Rao bound by means of measuring quantum coherence of the probe qubit.
We study the temperature sensing performance by the use of quantum signal-to-noise ratio (QSNR). It is indicated that there is an optimal encoding time that the QSNR can reach its maximum in the full-temperature regime.  In particular,  we find that  the QSNR reaches a finite upper bound in the weak coupling regime even when the temperature is close to absolute zero, which implies that
the sensing-error-divergence problem is avoided in our scheme. Our work opens a way for quantum sensing of temperature close to absolute zero in the BEC.

\end{abstract}


\maketitle
\section{\label{Sec:1}Introduction}
Temperature is the most basic physical quantity in both classical and quantum thermodynamics. Precise sensing of temperature is of wide importance for both fundamental nature science and the developing quantum technologies~\cite{Giazotto2006,Sanpera2019}. Traditional temperature measurement techniques such as time-of-flight absorption can be precise, but are often inherently destructive~\cite{Leanhardt2003,Hemmerling2006,Gati2006,Olf2015}. Using a small quantum system such as a two-level system ~\cite{Brunelli2011,Brunelli2012,White2014,Jevtic2015,Seveso2018,Razavian2019,Mitchison2020} or a harmonic oscillator~\cite{Mehboudi2019,Khan2022-,Khan2021}, acting as a quantum thermometer, to measure the temperature of a quantum reservoir has attracted much attention~\cite{Marzolino2013,Dragan2013,Mehboudi2015,Hohmann2016,Johnson2016,Seah2019,Tamascelli2020,Mancino2020,Kirkova2021,Rubio2021,Oghittu2022-,Adam2022-}. The back actions on the sample induced by the small quantum system are negligible and therefore the measurement process can be considered to be non-destructive. In quantum thermometry, one aims to enhance the temperature sensing precision as much as possible by using quantum features such as quantum coherence~\cite{Razavian2019,Mitchison2020,Stace2010,Candeloro2021}, quantum correlation~\cite{Francesca2020,Ather2021,Kenfack2021,Planella2022}, quantum Non-Markovian~\cite{zhou2021,Wu2021,Zhang2022}, coupling strength ~\cite{Mitchison2020,Correa2017}, periodic driving~\cite{Glatthard2022} and dynamic control~\cite{Feyles2019,Mukherjee2019,Kiilerich2018}.

Achieving ultra-low temperatures is essential for quantum simulation and computation in many experiment platforms~\cite{Bloch2008,Bloch2012,Sanpera2019}. With the development of cooling technology, cold atomic gases  have been successfully cooled to sub-nanokelvin regime, and even down to few picokelvins ~\cite{Leanhardt2003,Olf2015,Bloch2012}. However, there exist fundamental precision limitations that render accurately  measuring such low temperature remarkably difficult~\cite{Reeb2015,Correa2015,Paris2015,Hovhannisyan2018,Potts2019,Potts2020}. It has been shown for the thermal equilibrium probes, where the probes thermalize with the sample and the temperature sensing precision depends on the heat capacity, the sensing error will diverge exponentially as $T\rightarrow0$~\cite{Reeb2015,Correa2015,Paris2015}. Even though in some cases for non-thermal equilibrium probes, where quantum probes do not thermalize with the sample and reach a non-thermal steady state, instead of exponential divergence, a polynomial divergence still exist~\cite{Hovhannisyan2018,Potts2019,Potts2020}. Therefore, how to overcome the error-divergence in the low-temperature regime has become an open question. Recently, for a gapless harmonic oscillator probe some efforts have been paid to avoid the error-divergence~\cite{Hovhannisyan2018,Zhang2022,Glatthard2022}, but for a qubit probe it is still missing.

To address above question, we investigate the temperature sensing performance of a single-qubit quantum thermometer immersed in  a thermally equilibrated quasi-one-dimensional Bose-Einstein condensate (BEC). The qubit coupled to collective excitations of the BEC is described by a pure dephasing model~\cite{Recati2005,Cirone2009,Song2019}, which is exactly solvable~\cite{Breuer2007,Kuang1999,Tong1999} and is considered to be an ideal test bed for
investigating the problems of open quantum system such as the transitions from Markovian to non-Markovian dynamics~\cite{Haikka2011,Haikka2013, Yuan2017}.  In the sub-nK regime, we numerically investigate dynamical behaves of the  quantum signal-to-noise ratio (QSNR), which quantifies the ultimate sensitivity limit of the qubit thermometer. We find there is an optimal encoding time that the QSNR reaches its maximum. More interesting, we find the optimal encoding time is inversely proportional to temperature, which is helpful for determining the optimal encoding time when one employs the probe'nonequilibrium dynamics for low-temperature sensing~\cite{Hangleiter2015,Mancino2017,Hofer2017,Vasco2018,Bouton2020}. We further find that weaker coupling between the qubit and the BEC can enhance low-temperature sensing performance. In particular, as the coupling strength deceases, the optimal QSNR will increase to a same value for all temperatures. This indicates in our model there may be an upper bound of the QSNR in the weak-coupling limit. Under Ohmic spectrum density approximation, we obtain an analytical expression of the QSNR and successfully prove above numerical results. In the weak-coupling and low-temperature limits, we get a constant upper bound of the QSNR. This implies the relative error  does not diverge as temperature tends to absolute zero in our model. We will show that weak coupling, selecting optimal encoding time and sensitive of dephasing dynamics on the ultra-low temperature change  can be used to avoid the error-divergence in our model.

\section{\label{Sec:2}Physical model }
\begin{figure}
\includegraphics[width=0.99\columnwidth]{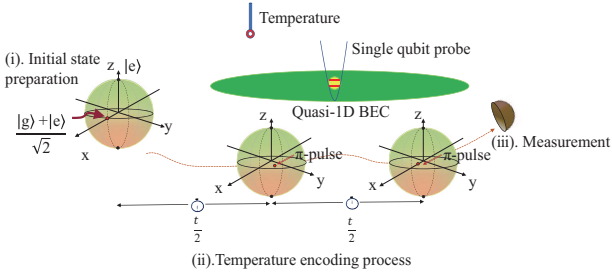}
\caption{(color online) Schematic of the physical system under
consideration: An  atomic qubit is immersed in a thermally equilibrated quasi-1D
BEC, which acts as a quantum sensor to estimate the temperature of the BEC. Our sensing protocol as following: (i) First we initialize the probe qubit in the superposition state $(|g\rangle +|e\rangle)/\sqrt{2}$, which is denoted by the Bloch sphere. (ii) Then the probe qubit  undergoes a temperature encoding process via interacting with the BEC. (iii) Finally,  choosing an optimal encoding time, we perform a measurement $\hat{\sigma}_{x}$ on the probe qubit, which can saturate the quantum Cram\'{e}r-Rao bound.} \label{model}
\end{figure}

As shown in Fig.~\ref{model}, we consider a single impurity-atomic qubit immersed in a thermally equilibrated quasi-1D BEC at temperature $T$. The qubit is confined in a harmonic trap $V_A({\bf x})=m_A\omega_{A}^{2}{\bf x}^{2}/2$ that is independent of the internal states, where $m_A$ is the mass of the impurity and $\omega_{A}$ is the trap frequency. For $\hbar\omega_{A}\gg k_{B}T$, the spatial wave function of the qubit is the ground state of $V_{A}(\mathbf{x})$, i.e., $\varphi_{A}({\bf x})=\pi^{-3/4}\ell_{A}^{-3/2}\exp[-{\bf x}^{2}/(2\ell_{A}^{2})]$ with $\ell_{A}=\sqrt{\hbar/(m_{A}\omega_{A})}$. The Hamiltonian of the qubit is $\hat{H}_{A}=\Omega_{A}|e\rangle\langle e|$, where $\Omega_{A}$ is level splitting between the ground ($|g\rangle$) and excited ($|e\rangle$) states.

For the BEC, we assume that the gas is  confined by a harmonic potential $V_{B}(\rho)=m_{B}\omega_{B}^{2}\rho^{2}/2$ in the x,y-directions($\rho^{2}=x^{2}+y^{2}$), where $\omega_{B}$ is the trap frequency. For sufficiently large $\omega_{B}$, the motion of the atoms along the $x$, $y$ axis is frozen to the ground state of $V_{B}(\rho)$, i.e., $\varphi_{B}(r)=\pi^{-1/2}\ell_{B}^{-1}\exp[-\rho^{2}/(2\ell_{B}^{2})]$ with $\ell_{B}=\sqrt{\hbar/(m_{B}\omega_{B})}$, which effectively reduces the BEC into a quasi-1D one. Finally, for small $T$, we may assume that most of the  atoms are condensed to zero momentum state with a line density  $n$. Then, following Bogoliubov's method, the uncondensed atoms are described by the quasiparticle Hamiltonian
\begin{equation}
\hat{H}_B =\sum_{{\bf k}\neq 0}\epsilon_{\mathbf k}\hat{b}^{\dag}_{{\bf k}}\hat{b}_{{\bf k}},\label{exc}
\end{equation}
where $\hat b_{\mathbf k}$ is annihilation operator for the quasiparticle with the wave vector ${\mathbf k}$.  Excitation energy $\epsilon_{\mathbf k}$ is
\begin{align}
\epsilon_{\mathbf k}=\sqrt{E_{\mathbf k}^{2}+2ng_{B}E_{\mathbf k}}\label{ek}
\end{align}
 where $E_{\mathbf k}=\hbar^{2}k^{2}/(2m_{B})$ is the kinetic energy and  $g_{B}=2\hbar^{2}a_{B}/(m_{B}\ell_{B}^{2})$ is coupling constant with $a_{B}$ being the $s$-wave scattering length. In this work, we take the Bogoliubov Excitations in Eq.~(\ref{exc}) as the reservoir of the  qubit sensor.

For the sensor-reservoir coupling, the qubit probe undergoes spin-dependent $s$-wave collisions with the ultracold gas. We assume the qubit and the gas interact only when the qubit is in state $|e\rangle$ ,which can be achieved by tuning the scattering length for state $|g\rangle$ to zero via Feshbach resonance~\cite{Chin2010}. Let $a_{AB}$ be the  scattering length in state $|e\rangle$, the qubit-BEC interaction Hamiltonian is then
\begin{equation}
\hat{H}_{AB}=\delta_{e}|e\rangle\langle e|+|e\rangle\langle e|
\sum_{\mathbf{k}\neq 0}g_{\mathbf k}\left(\hat{b}_{\mathbf{k}}+\hat{b}^{\dag}_{\mathbf{k}}\right),
\end{equation}
where $\delta_{e}=2\hbar^{2}na_{AB}/[m_{AB}(\ell^{2}_{A}+\ell^{2}_{B})]$ is the excited level shift due to the collision, $m_{AB}=m_{A}m_{B}/(m_{A}+m_{B})$ is the reduced mass, and the sensor-reservoir coupling parameter is
\begin{equation}
g_{\mathbf k}=\frac{\delta_{e}}{\sqrt{nL}}\sqrt{\frac{E_{\mathbf k}}{\epsilon_{\mathbf k}}}e^{-(\ell_{A}k)^{2}/4}\label{gk}
\end{equation}
with $L$ being the size of the quasi-1D BEC .

Now the total Hamiltonian, $\hat H=\hat H_{A}+\hat H_{B}+\hat H_{AB}$, is
\begin{equation}
\hat{H}=(\Omega_{A}+\delta_{e})|e\rangle\langle
e|+\sum_{{\mathbf k}\neq 0}\epsilon_{\mathbf k}\hat{b}^{\dag}_{\mathbf k}\hat{b}_{\mathbf k}+|e\rangle\langle e| \sum_{\mathbf{k}\neq0}g_{\mathbf k}\left(\hat{b}_{\mathbf{k}}+\hat{b}^{\dag}_{\mathbf{k}}\right),\label{hami}
\end{equation}
Since $\hat H_{A}$ commutes with $\hat H_{AB}$, the dynamics of the impurity qubit in reservoir is purely dephasing.

\section{\label{Sec:3} Temperature sensing protocol by a dephasing qubit }
As shown in Fig.~\ref{model}, we consider the following temperature sensing protocol. (i) First we initialize the probe qubit in the superposition state $(|g\rangle +|e\rangle)/\sqrt{2}$, which is denoted by the Bloch sphere. (ii) Then the Probe qubit  undergoes a temperature encoding process via interacting with the BEC. (iii) Finally,  choosing an optimal encoding time, we perform a  measurement $\hat{\sigma}_{x}$ on the probe qubit, which can saturate the quantum Cram\'{e}r-Rao bound. Next we will specify the sensing protocol. The initial state of the total system is prepared as
$(|g\rangle +|e\rangle)/\sqrt{2}\otimes\hat{\rho}_{B}(T)$,  where $\hat{\rho}_{B}(T)=\prod_{\mathbf k}\left(1-e^{\beta\epsilon_{\mathbf k}}\right)e^{-\beta\epsilon_{\mathbf k}b_{{\mathbf k}}^{\dag }b_{{\bf k}}}$ is  a thermal state of the BEC, where $\beta=1/k_{B}T$ with $k_{B}$ and $T$ being the Boltzmann constant and temperature, respectively. Let the whole system evolve under the control of the Hamiltonian (\ref{hami}) for a certain time $t/2$, after which a $\pi$-pulse about $x$ is applied to the qubit. Then the system is allowed to evolve for the same time period $t/2$ and another $\pi$-pulse is applied. Through these processes, the quantum state of the qubit probe at time $t$ can be given as
\begin{equation}
\rho_{A}(t)=\frac{1}{2}\left(
                \begin{array}{cc}
                  1 &e^{-\Gamma(t,T)} \\
                  e^{-\Gamma(t,T)} & 1 \\
                \end{array}
              \right)
,\label{state}
\end{equation}
where $\Gamma(t,T)$ is the dephasing factor with following expression
\begin{equation}
\Gamma(t,T)=\sum_{\mathbf{k}\neq0}\frac{2g_{\mathbf k}^{2}}{\epsilon_{\mathbf k}^{2}}\sin\left(\frac{\epsilon_{\mathbf k}t}{2\hbar}\right)^{2}\coth\left(\frac{\epsilon_{\mathbf k}}{2k_{B}T}\right).\label{gami}
\end{equation}
From above equation, one see the temperature information are encoded into the sensor'dephasing factor after encoding time $t$. Substituting $g_{\mathbf k}$  in Eq.~(\ref{gk}) into above equation and using the continuum limit $\frac{1}{L}\sum_{\mathbf{k}}\rightarrow\frac{1}{\pi}\int^{\infty}_{0}dk$, we further obtain the dephasing factor as
\begin{equation}
\Gamma(t,T)=P\int^{\infty}_{0}dk\frac{e^{-(\ell_{A}k)^{2}/2}}{\epsilon_{\mathbf k}(E_{\mathbf k}+2ng_{B})}\sin\left(\frac{\epsilon_{\mathbf k}t}{2\hbar}\right)^{2}\coth\left(\frac{\epsilon_{\mathbf k}}{2k_{B}T}\right)\label{gam2}
\end{equation}
with the parameter $P=8n\hbar^{4}a_{AB}^{2}/[\pi m_{AB}^{2}\left(\ell^{2}_{A}+\ell^{2}_{B}\right)^{2}]$.

In the following, we introduce the quantum parameter estimate theory to quantify temperature sensing precision of the state in Eq.~(\ref{state}). As is well-known, the temperature sensing precision is restricted to the  quantum Cram\'{e}r-Rao bound
 \begin{equation}
\delta T \geq \delta T_{QCR} \equiv\frac{1 }{\sqrt{\nu  \mathcal{F}^{Q}_{T}}}.\label{QCR}
\end{equation}
 Here $\delta T$ is the mean square error, $\nu$ represents the number of repeated experiments and $\mathcal{F}^{Q}_{T}$ denotes quantum Fisher information (QFI) with respect to the temperature $T$. It is more convenient for a qubit to express QFI in Bloch representation. Any qubit state in the Bloch sphere representation can be written as $\hat{\rho}=1/2(\hat{I}+\mathbf{w}\cdot\mathbf{\hat{\sigma}})$, where $\hat{I}$ is $2$ identity matrix, $\mathbf{w}=(w_{x},w_{y},w_{z})^{T}$ is the real Bloch vector and $\mathbf{\hat{\sigma}}=(\hat{\sigma}_{x},\hat{\sigma}_{y},\hat{\sigma}_{z})$ represents the pauli matrices. The eigenvalues of the density operator $\rho$ can be given as $(1\pm w)/2$, where $w$ is the length
of the Bloch vector. The length $w=1$ for pure state and $w<1$ for mixed state. In the  Bloch sphere representation the QFI with respect to the estimated parameter $\lambda$ can be given as follows \cite{Wei2013, Jing2019}
\begin{equation}\label{fisher1}
 \mathcal{F}^{Q}_{\lambda}=\left\{
                         \begin{array}{ll}

                            \left|\partial_{\lambda}\mathbf{w}\right|^{2}, & \hbox{$w=1;$} \\
                           \left|\partial_{\lambda}\mathbf{w}\right|^{2}+\frac{\left(\mathbf{w}\hspace{0.05cm}\cdot \hspace{0.05cm}\partial_{\lambda}\mathbf{w}\right)^{2}}{1-w^{2}}, & \hbox{$w < 1$.}

                         \end{array}
                       \right.,
\end{equation}
where $\partial_{\lambda}$ denotes the derivative with respect to the estimated parameter $\lambda$. Therefore, to obtain the $\mathcal{F}^{Q}_{T}$ of the quantum  state in Eq. (\ref{state}) , we rewritten the quantum  state as $\rho_{A}(t)=1/2(I+\mathbf{w}\cdot\mathbf{\sigma})$, where the Bloch vector $\mathbf{w}=(e^{-\Gamma},e^{-\Gamma},0)$. In the light of above equation, the concrete expression of the $\mathcal{F}^{Q}_{T}$ is obtained easily as follows
\begin{equation}\label{fisher2}
 \mathcal{F}^{Q}_{T}=\frac{\left(\partial_{T}\Gamma\right)^{2}}{e^{2\Gamma}-1}.
\end{equation}

The temperature sensing performance can be characterized by the QSNR
 \begin{equation}
\mathcal{Q}_{T}=T^{2}\mathcal{F}^{Q}_{T}.\label{QSNR}
\end{equation}
From the QCR bound in Eq.~(\ref{QCR}), the optimal relative error $(\delta T)_{min}/T$ and the QSNR $\mathcal{Q}_{T}$ has the following relation
\begin{equation}
\frac{(\delta T)_{min}}{T}=\frac{1}{\sqrt{\nu\mathcal{Q}_{T}}},
\end{equation}
which means that the larger is the QSNR the better is the temperature sensing performance.

The last step of the temperature sensing protocol, we propose a  measurement scheme at time $t$ which can saturate the quantum Cram\'{e}r-Rao bound. For a two-level system, the Fisher information associated with the measurement can be presented as~\cite{Mitchison2020}
\begin{equation}\label{fisher3}
\mathcal{F}_{T}=\frac{1}{\langle\Delta \hat{X}^{2}\rangle}\left(\frac{\partial\langle\hat{X}\rangle}{\partial T}\right)^{2},
\end{equation}
where $\langle\hat{X}\rangle$ and $\langle\Delta \hat{X}^{2}\rangle$ are mean and variance of the measured observable. The QFI is the upper bound of the Fisher information associated with the measurement $\hat{X}$, i.e.,  $\mathcal{F}^{Q}_{T}=\max_{\hat{X}}\mathcal{F}_{T}(\hat{X})=\mathcal{F}_{T}(\hat{\Lambda})$. It is very important to find an optimal measurement $\hat{\Lambda}$ for experimental implementation. In this work we choose $\langle\hat{\sigma}_{x}\rangle$ as the measurement signal, which can be observed using Ramsey interferometry~\cite{Adam2022-,Scelle2013,Cetina2016}.

Based on the quantum state of the single-atom temperature sensor given by Eq. (\ref{state}), it is straightforward to obtain
\begin{eqnarray}
\langle\hat{\sigma}_{x}\rangle=e^{-\Gamma},\hspace{0.2cm}
\langle\Delta\hat{\sigma}_{x}^{2}\rangle=1-e^{-2\Gamma}.
\label{sig}
\end{eqnarray}
The Fisher information associated with the measurement $\hat{\sigma}_{x}$ is obtained by  substituting above equation into Eq.~(\ref{fisher3})
\begin{equation}\label{fisher4}
 \mathcal{F}_{T}(\sigma_{x})=\frac{\left(\partial_{T}\Gamma\right)^{2}}{e^{2\Gamma}-1},
\end{equation}
which is exactly eqaul to  the quantum Fisher  information given by Eq. (11). Therefore, we can conclude that the sensitivity of the temperature sensor in present scheme can saturate the quantum Cram\'{e}r-Rao bound  through performing the  measurement $\hat{\sigma}_{x}$ on the probe qubit

\section{\label{Sec:4}Numerical results of the temperature sensing performance}

In this section we investigate numerically the temperature sensing performance via studying the dynamics of QSNR.
For facilitating the procedure of numerical calculation, we transfer the parameters in the dephasing factor in Eq.~(\ref{gam2}) into dimensionless form as
\begin{equation}
\Gamma(\tilde{t},\tilde{T})=\tilde{P}\int^{\infty}_{0}d\tilde{k}\frac{e^{\frac{-(\sigma\tilde{k})^{2}}{2}}}{\tilde{\epsilon}_{\mathbf k}(\tilde{k}^{2}+2\alpha)}\sin\left(\frac{\tilde{\epsilon}_{\mathbf k}\tilde{t}}{2\tilde{T}}\right)^{2}\coth\left(\frac{\tilde{\epsilon}_{\mathbf k}}{2\tilde{T}}\right),\label{gam3}
\end{equation}
where the dimensionless parameter $\tilde{P}=32na_{AB}^{2}(m_{A}+m_{B})^{2}/\left[\pi\ell_{B}m_{A}^{2}(1+\sigma^{2})^{2}\right]$
with $\sigma=\ell_{A}/\ell_{B}$, the dimensionless wave vector $\tilde{k}=\ell_{B}k$,  excitation energy $\tilde{\epsilon}_{\mathbf k}=\sqrt{\tilde{k}^4+2\alpha\tilde{k}^2}$ with $\alpha=4na_{B}$ ,time $\tilde{t}=\omega_{T}t$ with $\omega_{T}=k_{B}T/\hbar$ and temperature $\tilde{T}=2k_{B}T/(\hbar\omega_{B})$. To present our results, we consider a single $^{23}$Na atom with the mass $m_{A}\approx3.82\times10^{-26}$kg  immersed in a $^{87}$Rb BEC with the atomic mass $m_{B}\approx14.45\times10^{-26}$kg. We assume a typical trap frequency $\omega_{B}=2\pi\times 10^{3}\,{\rm Hz}$, the corresponding harmonic oscillator width is $\ell_{B}\simeq 3.4\times 10^{-5}\,{\rm cm}$. Next, we consider a typical condensate peak density of $10^{14}\,{\rm cm}^{-3}$, the line density is then $n=3.6\times 10^{5}\,{\rm cm}^{-1}$. In this work, the
 scattering length of the BEC takes its nature value of $a_{B}=a_{Rb}\approx5.3$nm and the dimensionless parameter $\sigma$ is given by $\sigma=0.5$. When it is not a variable, we take the scattering length $a_{AB}=55a_{0}\approx2.9$nm with $a_{0}$ being the Bohr radius\cite{Haikka2011}.  Consequently, we have $\tilde{P}\approx 0.13$ and $\alpha\approx0.77$.
\begin{figure}
\includegraphics[width=0.75\columnwidth]{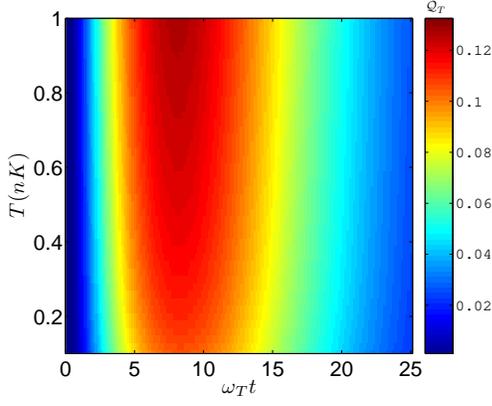}
\caption{(color online) The QSNR as a function of dimensionless time $\omega_{T}t$ and temperature $T$.The values of relevant parameters are shown in the first paragraph of  section \ref{Sec:4}.} \label{QSNR3D}
\end{figure}

\begin{figure}[tbp]
\includegraphics[width=0.99\columnwidth]{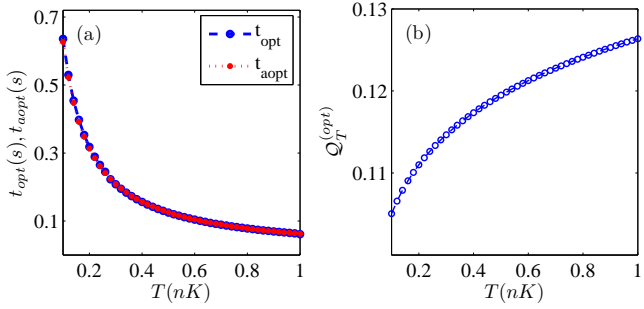}
\caption{(Color online) (a) Comparision of the numerical optimal encoding time $t_{opt}$ and the approximate optimal encoding time $t_{aopt}=8.2/\omega_{T}$. (b) The optimal QSNR  $\mathcal{Q}^{(opt)}_{T} $ as a function of the temperature $T$.} \label{maxQSNR}
\end{figure}

Figure~\ref{QSNR3D} plots the dynamical behaviors of the QSNR in the sub-nK range from $0.1$ nK to $1$ nK. It is shown the QSNR increases firstly and then decreases with time for full temperature range. This means that for a given temperature there is an optimal encoding time $t_{opt}$ that the QSNR reaches its maximum, which is called optimal QSNR $\mathcal{Q}^{(opt)}_{T}$. Moreover, figure~\ref{QSNR3D} reveals that the optimal encoding time $t_{opt}$  may satisfy the following relation
\begin{equation}
\omega_{T}t_{opt}= R\label{topt},
\end{equation}
where $R$ is a temperature-independent parameter. It follows that the optimal encoding time is inversely proportional to the temperature. To confirm this,  Figure~\ref{maxQSNR}(a) compares the numerical optimal encoding time $t_{opt}$ and the approximate optimal encoding time $t_{aopt}=8.2/\omega_{T}$.  As can be seen, the agreement is remarkable. Figure~\ref{maxQSNR}(b) corresponds to the maximal QSNR  $\mathcal{Q}^{(opt)}_{T} $ as a function of the temperature $T$. It is a natural result that the optimal QSNR increases with temperature.
\begin{figure}[tbp]
\includegraphics[width=0.75\columnwidth]{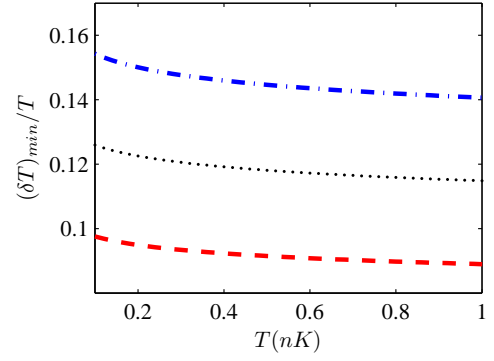}
\caption{(Color online) The optimal relative error $(\delta T)_{min}/T$ as a function of temperature $T$ after 400 (blue, dash-dotted), 600(black, dotted) and 1000(red, dashed) measurements.} \label{relerror}
\end{figure}

Figure~\ref{relerror} plots the optimal relative error $(\delta T)_{min}/T$ as a function of temperature $T$ after 400 (blue, dash-dotted), 600(black, dotted) and 1000(red, dashed) measurements. We can see in the sub-nk regime the optimal relative error is insensitive to temperature changes, which is different from the sensing schemes of the sub-nK temperature being encoded into the qubit's relative phase~\cite{White2014} or the harmonic oscillator's position quadrature~\cite{Mehboudi2019,Khan2022-}. In these schemes, as temperature decreases from $0.3$ nK, the upward tendency of the optimal relative error will become apparent. This implies our scheme may be more suitable for designing sub-nK-lower quantum sensor. From Fig.~(\ref{relerror}), one can see with only 400 measurements, the optimal relative error can be kept below $16\%$ for full temperatures( see blue dash-dotted line ).  As the number of the measurements increases to  $1000$ , the optimal relative error will drop below $10\%$ (see red dashed line). For the temperature sensing performance, the qubit sensor is comparable to the harmonic oscillator sensor in the sub-nk regime~\cite{Mehboudi2019,Khan2022-}.

\begin{figure}[tbp]
\includegraphics[width=0.99\columnwidth]{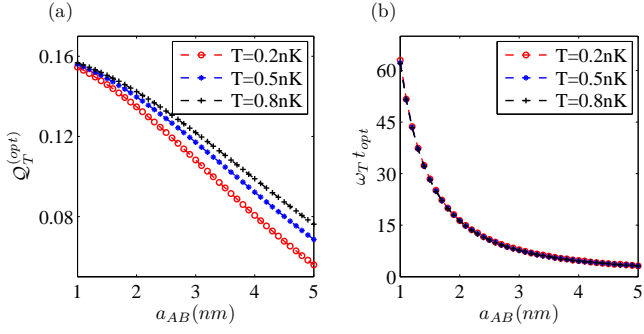}
\caption{(Color online) (a)  The optimal QSNR  $\mathcal{Q}^{(opt)}_{T}$ as a function of the scattering length $a_{_{AB}}$ for $0.2 nK$, $0.5 nK$ and $0.8 nK$. (b) The dimensionless optimal encoding time $\omega_{T}t_{opt}$ as a function of the scattering length $a_{_{AB}}$ for  $0.2 nK$, $0.5 nK$ and $0.8 nK$. Other related parameters are the same as  in Fig. \ref{QSNR3D}. } \label{aABQSNR}
\end{figure}

In the following, we study the effects of the coupling strength between the sensor and the BEC on the optimal QSNR and the optimal encoding time.
From Eq.~(\ref{gk}), we see the coupling strength can be well controlled by adjusting the scattering length $a_{_{AB}}$ via Feshbach resonance~\cite{Chin2010}. Therefore, figure~\ref{aABQSNR}(a) plots the optimal QSNR  $\mathcal{Q}^{(opt)}_{T}$ as a function of the scattering length $a_{_{AB}}$ for $0.2 nK$, $0.5 nK$ and $0.8 nK$. It is shown the optimal QSNR $\mathcal{Q}^{(opt)}_{T}$ decreases as increases the scattering length  $a_{_{AB}}$ for all temperatures. It is demonstrated that weaker coupling can enhance temperature sensing precision, which also holds in the temperature sensing process of a dephasing qubit immersed in a cold Fermi gas~\cite{Mitchison2020}, but is contrary for a harmonic oscillator probe, where strong coupling is considered as a resource to enhance the temperature sensing precision~\cite{Correa2017}. More interestingly, as the scattering length $a_{_{AB}}$ decreases to $1$nm, all optimal QSNRs  increase to the same value. This implies there may be a temperature-independent upper limit value of the QSNR when the scattering length  $a_{_{AB}}$ is small enough. Corresponding, the optimal encoding time $\omega_{T}t_{opt}$ is shown in Fig.~\ref{aABQSNR}(b). As can be seen, the optimal time $\omega_{T}t_{opt}$  increases with decreaseing the scattering length  $a_{_{AB}}$.  Moreover, the three curves for $0.2 nK$, $0.5 nK$ and $0.8 nK$ overlap very well, which indicates the equation~(\ref{topt}) still holds for different coupling strengths and only the temperature-independent parameter $R$ increases with decreasing the scattering length  $a_{_{AB}}$. Combining Fig.~\ref{aABQSNR}(a) and Fig. ~\ref{aABQSNR}(b), we see there is a trade-off between  optimal QSNR and optimal encoding time, which is controlled by  the scattering length  $a_{_{AB}}$. The price of obtaining larger optimal QSNR has to pay longer optimal encoding time.
\begin{figure}[tbp]
\includegraphics[width=0.75\columnwidth]{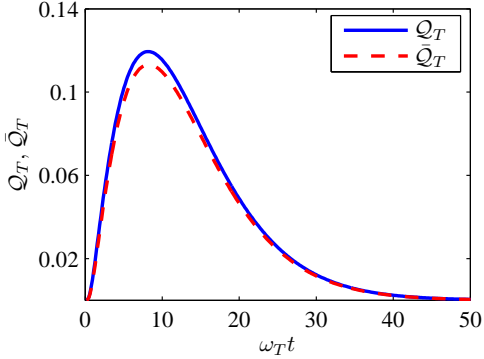}
\caption{(Color online) Comparison the time evolution of the numerical QSNR $\mathcal{Q}_{T}$ and the analytical QSNR $\bar{\mathcal{Q}}_{T}$ in Eq.(\ref{anaq}). The temperature $T=0.5nK$, other related parameters are the same as  in Fig. \ref{QSNR3D}. } \label{compareq}
\end{figure}

\section{\label{Sec:5} Analytical results of the temperature sensing performance}

In this section, we study analytically the temperature sensing performance of the dephasing qubit and try to give an analysis of above numerical results under some approximations. For the wave vector $k\ll\xi^{-1}=\sqrt{m_{B}ng_{B}}/\hbar$ with $\xi$ being the healing length, the excitations of the BEC are phonons with the dispersion relation $\epsilon_{\mathbf k}=\hbar c_{s} k$, where $c_{s}=\sqrt{ng_{B}/m_{B}}$ is the velocity of sound.  We notice the phonon excitations make the major contributions for the time evolution of the QSNR. Therefore, as an approximation, in the full wave vector region,  we substitute the phonon dispersion relation $\epsilon_{\mathbf k}=\hbar c_{s} k$ into the sensor-reservoir coupling parameter $g_{\mathbf k}$ in Eq.~(\ref{gk}) and obtain an analytical expression of reservoir spectrum density function $J(\omega)\equiv\sum_{\mathbf k}|g_{\mathbf k}|^{2}\delta(\omega-\omega_{\mathbf k})$ as
\begin{equation}
J(\omega)=\eta\omega e^{-\frac{\omega^{2} }{\omega _{c}^{2}}}\label{j1},
\end{equation}
where the dimensionless reservoir coupling parameter
\begin{equation}
\eta=\frac{na_{AB}^{2}\ell_{B}^{3}(m_{A}+m_{B})^{2}}{\sqrt{2}\pi
(\ell_{A}^{2}+\ell_{B}^{2})^{2}m_{A}^{2}}\left(\frac{1}{na_{B}}\right) ^{\frac{3}{
2}}\label{eta}
\end{equation}
and the cutoff frequency $\omega _{c}=\sqrt{2}c_{s}/\ell_{A}$. From Eq.~(\ref{eta}), we see the reservoir coupling parameter $\eta$ is proportional to the square of the scattering length $a_{AB}$ . It is worth noting that the scattering length $a_{AB} $ only appears in the  dimensionless reservoir coupling parameter  $\eta$ in our model. So adjusting the  scattering length $a_{AB}$ is equivalent to controlling the reservoir coupling parameter  $\eta$. For $1$ nm $\leq a_{AB} \leq 5$nm, the reservoir coupling parameter $\eta$ is in the range of  $0.004\leq\eta\leq0.1$ with related parameters being the same as  in Fig. \ref{QSNR3D}.   For getting a concrete analytical expression of the QSNR, we further approximate the spectrum density function as the standard Ohmic spectrum density $J(\omega)=\eta\omega e^{-\frac{\omega }{\omega _{c}}}$~\cite{Benedetti2018,Sehdaran2019,Tan2022}. Then under condition $\omega_{c}/\omega_{T}>>1$ (in our model, $\omega_{c}/\omega_{T}=84$ for $T=1nK$ ), the dephasing factor
is given as
\begin{eqnarray}
\bar{\Gamma}
=\frac{\eta}{2}\ln (1+\omega _{c}^{2}t^{2})+\eta\ln \left[\frac{1}{\pi\omega
_{T}t}\sinh(\pi\omega_{T}t)\right].
\end{eqnarray}
Substituting above equation into Eqs.~(\ref{fisher2}) and (\ref{QSNR}),  we find QSNR analytically that
\begin{eqnarray}
\bar{\mathcal{Q}}_{T}=\frac{\eta^{2}\left[\pi\omega _{T}t\coth (\pi \omega _{T}t)-1\right] ^{2}}{
(1+\omega _{c}^{2}t^{2})^{\eta}\left[\frac{1}{\pi \omega _{T}t}\sinh (\pi
\omega _{T}t)\right] ^{2\eta}-1}. \label{anaq}
\end{eqnarray}
In Fig. (\ref{compareq}), we compare the time evolution of the numerical QSNR $\mathcal{Q}_{T}$ and the analytical QSNR $\bar{\mathcal{Q}}_{T}$ in Eq. (\ref{anaq}) with  $T=0.5$ nK. As can be seen, except for the slight difference at maximum point, the agreement is remarkable. In fact, such agreement still holds for other temperatures.

Now according to the analytical expression of QSNR $\bar{\mathcal{Q}}_{T}$, we would like to provide an analytical description of the numerical results obtained.   We firstly analyze the QSNR varying with time.   In order to find the extreme-value point $\bar{t}_{opt}$, we take the derivative of  QSNR $\bar{\mathcal{Q}}_{T}$ with respect time $t$. We shall analytically indicate  that there exists an optimal encoding time $\bar{t}_{opt}$.  When $t<\bar{t}_{opt}$, the derivative $\partial \bar{\mathcal{Q}}_{T}/\partial t >0$,  otherwise, $\partial \bar{\mathcal{Q}}_{T}/\partial t <0$. Therefore,  as shown in Fig. \ref{QSNR3D}, the QSNR increases firstly then decreases with time.
We further find, under the conditions of $\omega _{c}\bar{t}_{opt}>>1$ and $\pi\omega _{T}\bar{t}_{opt}>>1$ ( the conditions are well satisfied, see the numerical results in Fig.~\ref{aABQSNR} (b)), the optimal encoding time   satisfies the following equation
\begin{eqnarray}
\omega _{T}\bar{t}_{opt}=\frac{1}{\pi\eta }\left(1-\frac{1}{\zeta _{T}^{2\eta}e^{2\pi\eta\omega _{T}\bar{t}_{opt}}}\right), \label{anaopt1}
\end{eqnarray}
where $\zeta _{T}=\omega _{c}/(2\pi \omega _{T})$.

In the following, we shall analyze the optimal encoding time $\bar{t}_{opt}$ as the function of the temperature and the scattering length $a_{AB} $.  We will analytically  indicate why $\omega _{T}t_{opt}$ can be approximated as a temperature-independent parameter $R$ in Eq. (\ref{topt}) and whether there is  a concrete expression of $R$, which matches the behavior in Fig.~\ref{aABQSNR} (b). We rewrite Eq. (23) as the following form
\begin{eqnarray}
\omega _{T}\bar{t}_{opt}=\frac{1}{\pi\eta }\left(1-z-2z^{2}+o(z^{3})\right), \label{anaopt2}
\end{eqnarray}
where $z$ is a temperature-dependent parameter with following form
 \begin{eqnarray}
z(T)=\frac{1}{e^{2}}\left(\frac{2\pi k_{B}T}{\hbar\omega_{c}}\right)^{2\eta}. \label{z}
\end{eqnarray}
Above equation shows the parameter $z$ is an increasing function of temperature $T$ and decreasing function of reservoir coupling parameter $\eta$ due to $2\pi k_{B}T/\hbar\omega_{c}<1$. Substituting the values of related parameters in the first paragraph of  section \ref{Sec:4} into Eq. (~\ref{z}), we obtain  $0.05\leq z\leq0.13$ in the temperature range of $0.1nK\leq T \leq 1nK$ and $1 nm\leq a_{AB} \leq 5nm$.
 The temperature-dependent parameter $z$ can be further made Taylor series expansion of temperature at temperature $T_{0}$ as $z=z(T_{0})+2\eta(T-T_{0})/T_{0}+o(\eta^{2})$. In the sub-nK regime, we take $T_{0}=0.5 nK$, then obtain $|(T-T_{0})/T_{0}|\leq1$ and $z(T_{0})=1/(e^{2}27^{2\eta})$. Due to $\eta$ being a small quantity  and $z$ itself also a small quantity in Eq.~(\ref{anaopt2}), it is  reasonable to substitute $z\approx z(T_{0})=1/(e^{2}27^{2\eta})$ into Eq.~(\ref{anaopt2}) and obtain
\begin{eqnarray}
\omega _{T}\bar{t}_{opt}=\bar{R} =\frac{1}{\pi\eta }\left[1-z(T_{0})-2z^2(T_{0})+o\left(z^3(T_{0})\right)\right]. \label{anaopt3}
\end{eqnarray}

\begin{figure}[tbp]
\includegraphics[width=0.75\columnwidth]{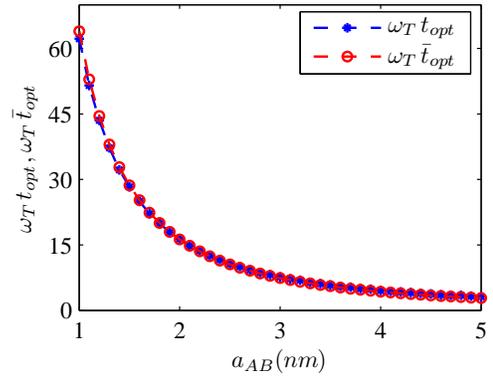}
\caption{(Color online) Comparison the numerical optimal encoding time $\omega _{T}t_{opt}$ and the analytical optimal encoding time $\omega _{T}\bar{t}_{opt}$ in Eq.~(\ref{anaopt3}) as a function of the scattering length $a_{_{AB}}$.} \label{comoptime}
\end{figure}

In Fig.~\ref{comoptime} we make a comparison between the numerical optimal encoding time $\omega _{T}t_{opt}$ and the analytical optimal encoding time $\omega _{T}\bar{t}_{opt}$ in Eq.~(\ref{anaopt3}) as a function of the scattering length $a_{_{AB}}$. Figure.~\ref{comoptime} indicates the good agreement between the numerical and analytical optimal encoding time and $R=\bar{R}$.

We now analytically investigate the influence of  the temperature and the scattering length $a_{AB}$ on   the optimal QSNR. Substituting  the analytically optimal encoding time given by Eq.~(\ref{anaopt2}) into Eq.~(\ref{anaq}), we can obtain the analytically optimal QSNR $\bar{\mathcal{Q}}^{(opt)}_{T}$  with the following expression
\begin{eqnarray}
\bar{\mathcal{Q}}^{(opt)}_{T}= (1-2\eta)z+(1-4\eta)z^{2}+2(1-6\eta)z^{3}+o(z^{4}). \label{aqmax}
\end{eqnarray}

\begin{figure}[tbp]
\includegraphics[width=0.99\columnwidth]{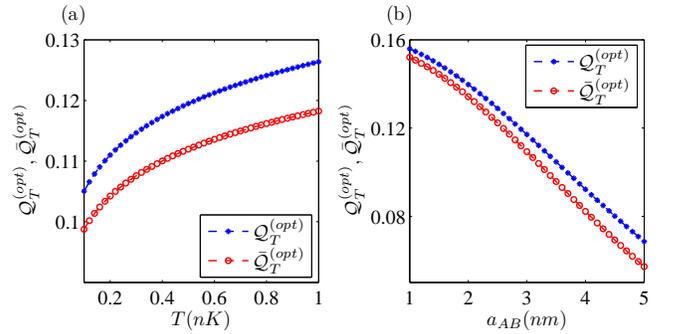}
\caption{(Color online) (a) Comparison the numerical optimal QSNR $\mathcal{Q}^{(opt)}_{T}$ and the analytical optimal QSNR  $\bar{\mathcal{Q}}^{(opt)}_{T}$ in Eq.~(\ref{aqmax})  as a function of temperature. the scattering length $a_{AB}$ is taken as $2.9 nm$.
 (b) Comparison the numerical optimal QSNR $\mathcal{Q}^{(opt)}_{T}$ and the analytical optimal QSNR  $\bar{\mathcal{Q}}^{(opt)}_{T}$ in Eq.~(\ref{aqmax})  as a function of scattering length $a_{AB}$. the temperature is taken as $T=0.5 nK$.} \label{comqmax}
\end{figure}

In order to compare the analytical and numerical results of  the optimal QSNR,  we have plotted the analytically and numerically optimal QSNR  as a function of temperature and the scattering length in Fig.~\ref{comqmax}(a) and Fig.~\ref{comqmax}(b), respectively.
From Fig.~\ref{comqmax}(a) and  Fig.~\ref{comqmax}(b), we can see that the analytical optimal QSNR is  highly consistent with the  numerical optimal QSNR.  In particular, the lower the temperature is or the smaller the scattering length is, the better the consistency between the analytical and numerical results is.

\begin{figure}[tbp]
\includegraphics[width=0.75\columnwidth]{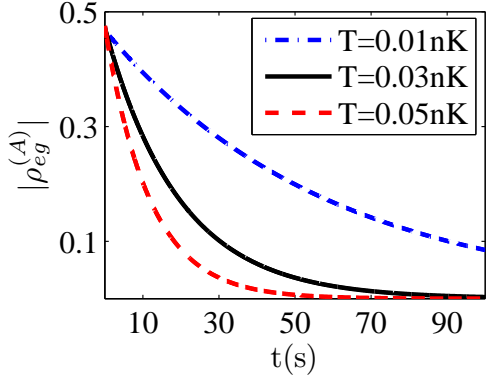}
\caption{(Color online) The dynamics of  quantum coherence $|\rho^{(A)}_{eg}|$ of the sensor for $0.01nK$, $0.03nK$ and $0.05nK$. The dimensionless reservoir coupling parameter $\eta=0.004$.  } \label{coherence}
\end{figure}

We now analytically study  the optimal QSNR in the weak coupling limit and show that  the optimal QSNR  can tend to the same value which is independent of temperature  when the scattering length decreases to  the $1 nm$ scale.  Specifically,  the reservoir coupling parameter $\eta$ becomes very small, $\eta\approx 0.004$,  as the scattering length $a_{AB}=1 nm$.  In this case, the temperature-dependent parameter $z$ in Eq.~(\ref{z}) becomes very insensitive to temperature. In fact, when $\eta=0.004$, it is calculated as $z(T=1nK)=0.132$ and $z(T=0.1nK)=0.130$, which shows the temperature effects on the optimal QSNR in Eq.~(\ref{aqmax}) becomes negligible. Thanks to the parameter $z$ being a decreasing function of reservoir coupling parameter $\eta$, one can conclude from Eq.~(\ref{aqmax}) that the optimal QSNR
is a decreasing function of reservoir coupling parameter $\eta$. This means weaker coupling can enhance the temperature sensing precision as shown in Fig.~\ref{aABQSNR}(a). Inspiring by these analyses, based on Eq.~(\ref{aqmax}), we can get an upper bound of the QSNR as
\begin{eqnarray}
\mathcal{Q}_{T}^{up}= \lim_{\eta\rightarrow 0}\bar{\mathcal{Q}}_{T}^{(opt)}=\frac{1}{e^{2}}+\frac{1}{e^{4}}+\frac{2}{e^{6}}+o\left(\frac{1}{e^{8}}\right)\approx0.16, \label{limq}
\end{eqnarray}
which is applicable in the range of $T<<\hbar\omega_{c}/k_{B}$ and $\eta<<1$ for an Ohmic reservoir. It is worth noting for arbitrarily low temperature, as long as the $\eta$ is small enough,  the QSNR also can close to the  upper bound $\mathcal{Q}_{T}^{up}$. This can be understood from the dynamics of  quantum coherence of the temperature sensor. Figure~\ref{coherence} plots the dynamics of quantum coherence $|\rho^{(A)}_{eg}|$ for $0.01nK$, $0.03nK$ and $0.05nK$. It is demonstrated that the dynamics is still sensitive with sub-nK-lower temperature change, although the sensitivity takes longer to be prominent.

Finally, we analytically discuss the behaviors of optimal QSNR when  temperature is close to absolute zero. It is well known that for the thermal equilibrium probes, the QSNR satisfies $\mathcal{Q}_{T}\sim e^{-\Delta/(k_{B}T)}$ with $\Delta$ being energy gap, which indicates the QSNR will decay exponentially fast to zero in the limit $T\rightarrow 0$, and for  a non-thermal equilibrium harmonic oscillator probe, a quartic  scaling law $\mathcal{Q}_{T}\sim T^{4}$ can be achieved, which demonstrates QSNR still decays to zero in the limit $T\rightarrow 0$ ~\cite{Hovhannisyan2018,Potts2019,Potts2020}. In our model, according to Eqs.~(\ref{z}) and (\ref{aqmax}), we can obtain  the scaling law $\bar{\mathcal{Q}}_{T}^{(opt)}\sim T^{2\eta}$ in the limit $T\rightarrow 0$. This implies that when the parameter $\eta$ approaches to zero, one
can prevent the optimal QSNR from decaying to zero as the temperature is close to absolute zero. As a matter of fact, in the weak coupling limit, using the limit relation $\lim_{x,y\rightarrow 0}x^{y}=1$, we can obtain
\begin{eqnarray}
\lim_{\eta,T\rightarrow 0
 }\bar{\mathcal{Q}}_{T}^{(opt)}=\mathcal{Q}_{T}^{up} \label{limq2},
\end{eqnarray}
where the upper bound of the QSNR $\mathcal{Q}_{T}^{up}$ is given by Eq. (28). Therefore, we can conclude that the temperature sensing error in our model does not diverge as temperature is close to absolute zero in the weak coupling limit.  We should point out the weak coupling, selecting optimal encoding time to achieve the optimal QSNR, and sensitivity of dephasing dynamics on the ultra-low temperature change are responsible for avoiding the error-divergence in our model.

\section{\label{Sec:6} Conclusions}

In conclusion, we have studied the quantum sensing scheme of temperature close to absolute zero in a quasi-one-dimensional BEC.   We have demonstrated that the sensitivity of the temperature sensor can saturate the quantum Cram\'{e}r-Rao bound by means of measuring quantum coherence of the probe qubit, and weaker coupling between the probe qubit and the BEC can enhance the temperature sensing precision.
We have investigated numerically and analytically the temperature sensing performance. It has been indicated that there is an optimal encoding time in which the QSNR can reach its maximum in the full-temperature regime.  In particular,  it has been found  that  the QSNR can reach a finite temperature-independent upper bound under  the weak coupling condition even when the temperature is close to absolute zero. This implies that the sensing-error-divergence problem is avoided in our scheme. It has been shown that weak coupling, selecting optimal encoding time and sensitive dephasing dynamics on the ultra-low temperature change are responsible for avoiding the error-divergence. Our work opens a way for quantum sensing of temperature close to absolute zero in the BEC.


\acknowledgments
J. B.  Yuan was supported by  NSFC (No. 11905053),   Scientific Research Fund of Hunan Provincial Education Department of China under Grant (No. 21B0647) and Hunan Provincial Natural Science Foundation of China under Grant (No. 2018JJ3006).  Y. J.  Song was supported by  NSFC (No. 12205088). S. Q Tang was supported by Scientific Research Fund of Hunan Provincial Education Department of China under Grant (No. 22A0507). L. M. Kuang was supported by NSFC   (Nos. 12247105, 1217050862,12247105, and 11935006) and the science and technology innovation Program of Hunan Province (No. 2020RC4047).

\end{document}